%Paper: hep-th/9207117
%From: vipul@guinness.ias.edu (Vipul Periwal)
%Date: Fri, 31 Jul 92 13:38:06 EDT

%This paper is in plain tex and all macropackages are included.
%
\catcode`@=11
\expandafter\ifx\csname inp@t\endcsname\relax\let\inp@t=\input
\def\input#1 {\expandafter\ifx\csname #1IsLoaded\endcsname\relax
\inp@t#1%
\expandafter\def\csname #1IsLoaded\endcsname{(#1 was previously loaded)}
\else\message{\csname #1IsLoaded\endcsname}\fi}\fi
\catcode`@=12

%%
%%  Font definitions for Computer Modern (CM) Fonts
%%
\font\twelverm=cmr12			\font\twelvei=cmmi12
\font\twelvesy=cmsy10 scaled 1200	\font\twelveex=cmex10 scaled 1200
\font\twelvebf=cmbx12			\font\twelvesl=cmsl12
\font\twelvett=cmtt12			\font\twelveit=cmti12
\font\twelvesc=cmcsc10 scaled 1200	\font\twelvesf=cmss12
\skewchar\twelvei='177			\skewchar\twelvesy='60

%  Define \...point macros to change fonts and spacings consistently

\def\twelvepoint{\normalbaselineskip=12.4pt plus 0.1pt minus 0.1pt
  \abovedisplayskip 12.4pt plus 3pt minus 9pt
  \belowdisplayskip 12.4pt plus 3pt minus 9pt
  \abovedisplayshortskip 0pt plus 3pt
  \belowdisplayshortskip 7.2pt plus 3pt minus 4pt
  \smallskipamount=3.6pt plus1.2pt minus1.2pt
  \medskipamount=7.2pt plus2.4pt minus2.4pt
  \bigskipamount=14.4pt plus4.8pt minus4.8pt
  \def\rm{\fam0\twelverm}          \def\it{\fam\itfam\twelveit}%
  \def\sl{\fam\slfam\twelvesl}     \def\bf{\fam\bffam\twelvebf}%
  \def\mit{\fam 1}                 \def\cal{\fam 2}%
  \def\sc{\twelvesc}		   \def\tt{\twelvett}
  \def\sf{\twelvesf}
  \textfont0=\twelverm   \scriptfont0=\tenrm   \scriptscriptfont0=\sevenrm
  \textfont1=\twelvei    \scriptfont1=\teni    \scriptscriptfont1=\seveni
  \textfont2=\twelvesy   \scriptfont2=\tensy   \scriptscriptfont2=\sevensy
  \textfont3=\twelveex   \scriptfont3=\twelveex  \scriptscriptfont3=\twelveex
  \textfont\itfam=\twelveit
  \textfont\slfam=\twelvesl
  \textfont\bffam=\twelvebf \scriptfont\bffam=\tenbf
  \scriptscriptfont\bffam=\sevenbf
  \normalbaselines\rm}

%	tenpoint

%%
%%	Various internal macros
%%

\def\beginlinemode{\endmode
  \begingroup\parskip=0pt \obeylines\def\\{\par}\def\endmode{\par\endgroup}}
\def\beginparmode{\endmode
  \begingroup \def\endmode{\par\endgroup}}
\let\endmode=\par
{\obeylines\gdef\
{}}
\def\singlespace{\baselineskip=\normalbaselineskip}

\def\oneandahalfspace{\baselineskip=\normalbaselineskip
  \multiply\baselineskip by 3 \divide\baselineskip by 2}
\def\doublespace{\baselineskip=\normalbaselineskip \multiply\baselineskip by 2}

\newcount\firstpageno
\firstpageno=2
\footline={\ifnum\pageno<\firstpageno{\hfil}\else{\hfil\twelverm\folio\hfil}\fi}
\def\toppageno{\global\footline={\hfil}\global\headline
  ={\ifnum\pageno<\firstpageno{\hfil}\else{\hfil\twelverm\folio\hfil}\fi}}
\let\rawfootnote=\footnote		% We must set the footnote style
\def\footnote#1#2{{\rm\singlespace\parindent=0pt\parskip=0pt
  \rawfootnote{#1}{#2\hfill\vrule height 0pt depth 6pt width 0pt}}}
\def\raggedcenter{\leftskip=4em plus 12em \rightskip=\leftskip
  \parindent=0pt \parfillskip=0pt \spaceskip=.3333em \xspaceskip=.5em
  \pretolerance=9999 \tolerance=9999
  \hyphenpenalty=9999 \exhyphenpenalty=9999 }
\def\dateline{\rightline{\ifcase\month\or
  January\or February\or March\or April\or May\or June\or
  July\or August\or September\or October\or November\or December\fi
  \space\number\year}}
\def\received{\vskip 3pt plus 0.2fill
 \centerline{\sl (Received\space\ifcase\month\or
  January\or February\or March\or April\or May\or June\or
  July\or August\or September\or October\or November\or December\fi
  \qquad, \number\year)}}

%%
%%	Page layout, margins, font and spacing (feel free to change)
%%

\hsize=6.5truein
\hoffset=0pt
%%\hoffset=1truein
\vsize=8.9truein
\voffset=0pt
%%\voffset=1truein
\parskip=\medskipamount
\def\\{\cr}
\twelvepoint		% selects twelvepoint fonts (cf. \tenpoint)
\doublespace		% selects double spacing for main part of paper (cf.
			%	\singlespace, \oneandahalfspace)
\overfullrule=0pt	% delete the nasty little black boxes for overfull box

%%
%%	The user definitions for major parts of a paper (feel free to change)
%%

  %  "Draft", Timestamp

	% Preprint number at upper right of title page

\def\title			%  Title on title page
  {\null\vskip 3pt plus 0.2fill
   \beginlinemode \doublespace \raggedcenter \bf}

\def\author			%  Author(s) name(s)  on title page
  {\vskip 3pt plus 0.2fill \beginlinemode
   \singlespace \raggedcenter\sc}

\def\affil			% Affiliations (can intermix with \author)
  {\vskip 3pt plus 0.1fill \beginlinemode
   \oneandahalfspace \raggedcenter \sl}

\def\abstract			% Begin abstract
  {\vskip 3pt plus 0.3fill \beginparmode
   \baselineskip=16truept ABSTRACT: }

\def\endtitlepage		% End title page, begin body of paper
  {\endpage			% 	This subsumes \body
   \body}
\let\endtopmatter=\endtitlepage

\def\body			% Begin text body;  can be used to end
  {\beginparmode}		% \title, \author, \affil, \abstract,
				% \reference, or \figurecaption modes

\def\head#1{			% Head;  NOTE enclose the text in {}
  \goodbreak\vskip 0.5truein	%  e.g., \head{I. Introduction}
  {\immediate\write16{#1}
   \raggedcenter \uppercase{#1}\par}
   \nobreak\vskip 0.25truein\nobreak}

\def\beginitems{
\par\medskip\bgroup\def\i##1 {\item{##1}}\def\ii##1 {\itemitem{##1}}
\leftskip=36pt\parskip=0pt}
\def\enditems{\par\egroup}

\def\beneathrel#1\under#2{\mathrel{\mathop{#2}\limits_{#1}}}

\def\refto#1{$^{#1}$}		% For references in text as superscript

\def\references			% Begin references -- basic format is Phys Rev
  {\head{References}		% I.e., volume, page, year (space after commas).
   \beginparmode
   \frenchspacing \parindent=0pt \leftskip=1truecm
   \parskip=3pt plus 3pt \baselineskip=17truept %
   \everypar{\hangindent=\parindent}}

\gdef\refis#1{\item{#1.\ }}			% Ref list numbers.

\gdef\journal#1, #2, #3, 1#4#5#6{		% Journal reference.  Comma sets
    {\sl #1~}{\bf #2}, #3 (1#4#5#6)}		% off: name, vol, page, year

\def\endreferences{\body}

\def\figurecaptions		% Begin figure captions
  {\endpage
   \beginparmode
   \head{Figure Captions}
}

\def\endpage			%  Eject a page
  {\vfill\eject}

\def\endpaper			%  Ways to say goodbye
  {\endmode\vfill\supereject}

%%
%%	AmSTeX compatability definitions
%%
%%	To run a TeX file originally intended for AmSTeX, only small changes
%%	should be necessary (I hope).  Use the line \input jnl at the start.
%%	Remove the lines \input amstex, \documentstyle{itpjnl} at the
%%	beginning;  also remove all the page layout stuff (\parindent=1cm,
%%	\hsize=5.28125in etc.)  The page layout is now done automatically.
%%	Also OMIT the qualifier \magnification=1200 when you IMPRINT the
%%	.dvi file.  (\TagsOnRight is harmless, you can take it out or leave
%%	it in.)  I believe most AmSTeX will work with no change.  One problem
%%	is \footnote, which is a little different in that it now needs to
%%	have an explicit asterisk *  (or whatever) included, like this:
%%		\footnote*{Text winds up at bottom of page.}
%%	This is discussed on p. 116 of the TeXbook.  IGNORE the AmSTeX
%%	documentation (if you can call it that);  refer to the TeXbook.
%%
%%	Note that many commands in AmSTeX have their equivalents in the
%%	TeXbook, perhaps with different names and slightly differing
%%	usage. E.g., the old \align in AmSTeX is replaced by \eqalign
%%	(p. 190) and \aligntag is replaced by \eqalignno (p. 192).
%%	\align and \aligntag still work, but I recommend that you use
%%	\eqalign and \eqalignno in documents run under jnl.
%%
%%	See me if you have any problems  -- Doug.
%%

\def\heading				% Heading
  {\vskip 0.5truein plus 0.1truein	% e.g., \heading I. NOTES \endheading
   \beginparmode \def\\{\par} \parskip=0pt \singlespace \raggedcenter}

\def\subheading				% Subheading
  {\vskip 0.25truein plus 0.1truein	% e.g., \subheading{A. The Problem}
   \beginlinemode \singlespace \parskip=0pt \def\\{\par}\raggedcenter}

\def\tag#1$${\eqno(#1)$$}

\def\align#1$${\eqalign{#1}$$}

\def\aligntag#1$${\gdef\tag##1\\{&(##1)\cr}\eqalignno{#1\\}$$
  \gdef\tag##1$${\eqno(##1)$$}}

\def\endaligntag{}

\def\overset #1\to#2{{\mathop{#2}\limits^{#1}}}
\def\underset#1\to#2{{\let\next=#1\mathpalette\undersetpalette#2}}
\def\undersetpalette#1#2{\vtop{\baselineskip0pt
\ialign{$\mathsurround=0pt #1\hfil##\hfil$\crcr#2\crcr\next\crcr}}}

%%
%%	Various little user definitions
%%

\def\ref#1{Ref.~#1}			% 	for inline references
\def\Ref#1{Ref.~#1}			% 	ditto
\def\[#1]{[\cite{#1}]}
\def\cite#1{{#1}}
			% For figure numbers
		% For citation of equation numbers
	%	ditto
			%	ditto
			%	ditto
		%	ditto
\def\(#1){(\call{#1})}
\def\call#1{{#1}}
\def\taghead#1{}
\def\frac#1#2{{#1 \over #2}}

\def\12{{1\over2}}

\def\ie{{\it i.e.,\ }}

\def\sla{\raise.15ex\hbox{$/$}\kern-.57em}
\def\leaderfill{\leaders\hbox to 1em{\hss.\hss}\hfill}
\def\twiddle{\lower.9ex\rlap{$\kern-.1em\scriptstyle\sim$}}
\def\bigtwiddle{\lower1.ex\rlap{$\sim$}}
\def\gtwid{\mathrel{\raise.3ex\hbox{$>$\kern-.75em\lower1ex\hbox{$\sim$}}}}
\def\ltwid{\mathrel{\raise.3ex\hbox{$<$\kern-.75em\lower1ex\hbox{$\sim$}}}}
\def\square{\kern1pt\vbox{\hrule height 1.2pt\hbox{\vrule width 1.2pt\hskip 3pt
   \vbox{\vskip 6pt}\hskip 3pt\vrule width 0.6pt}\hrule height 0.6pt}\kern1pt}
\def\tdot#1{\mathord{\mathop{#1}\limits^{\kern2pt\ldots}}}

\def\pmb#1{\setbox0=\hbox{#1}%
  \kern-.025em\copy0\kern-\wd0
  \kern  .05em\copy0\kern-\wd0
  \kern-.025em\raise.0433em\box0 }

\catcode`@=11
\newcount\tagnumber\tagnumber=0

\immediate\newwrite\eqnfile
\newif\if@qnfile\@qnfilefalse
\def\write@qn#1{}
\def\writenew@qn#1{}
\def\w@rnwrite#1{\write@qn{#1}\message{#1}}
\def\@rrwrite#1{\write@qn{#1}\errmessage{#1}}

\def\taghead#1{\gdef\t@ghead{#1}\global\tagnumber=0}
\def\t@ghead{}

\expandafter\def\csname @qnnum-3\endcsname
  {{\t@ghead\advance\tagnumber by -3\relax\number\tagnumber}}
\expandafter\def\csname @qnnum-2\endcsname
  {{\t@ghead\advance\tagnumber by -2\relax\number\tagnumber}}
\expandafter\def\csname @qnnum-1\endcsname
  {{\t@ghead\advance\tagnumber by -1\relax\number\tagnumber}}
\expandafter\def\csname @qnnum0\endcsname
  {\t@ghead\number\tagnumber}
\expandafter\def\csname @qnnum+1\endcsname
  {{\t@ghead\advance\tagnumber by 1\relax\number\tagnumber}}
\expandafter\def\csname @qnnum+2\endcsname
  {{\t@ghead\advance\tagnumber by 2\relax\number\tagnumber}}
\expandafter\def\csname @qnnum+3\endcsname
  {{\t@ghead\advance\tagnumber by 3\relax\number\tagnumber}}

\def\equationfile{%
  \@qnfiletrue\immediate\openout\eqnfile=\jobname.eqn%
  \def\write@qn##1{\if@qnfile\immediate\write\eqnfile{##1}\fi}
  \def\writenew@qn##1{\if@qnfile\immediate\write\eqnfile
    {\noexpand\tag{##1} = (\t@ghead\number\tagnumber)}\fi}
}

\def\callall#1{\xdef#1##1{#1{\noexpand\call{##1}}}}
\def\call#1{\each@rg\callr@nge{#1}}

\def\each@rg#1#2{{\let\thecsname=#1\expandafter\first@rg#2,\end,}}
\def\first@rg#1,{\thecsname{#1}\apply@rg}
\def\apply@rg#1,{\ifx\end#1\let\next=\relax%
\else,\thecsname{#1}\let\next=\apply@rg\fi\next}

\def\callr@nge#1{\calldor@nge#1-\end-}
\def\callr@ngeat#1\end-{#1}
\def\calldor@nge#1-#2-{\ifx\end#2\@qneatspace#1 %
  \else\calll@@p{#1}{#2}\callr@ngeat\fi}
\def\calll@@p#1#2{\ifnum#1>#2{\@rrwrite{Equation range #1-#2\space is bad.}
\errhelp{If you call a series of equations by the notation M-N, then M and
N must be integers, and N must be greater than or equal to M.}}\else%
 {\count0=#1\count1=#2\advance\count1
by1\relax\expandafter\@qncall\the\count0,%
  \loop\advance\count0 by1\relax%
    \ifnum\count0<\count1,\expandafter\@qncall\the\count0,%
  \repeat}\fi}

\def\@qneatspace#1#2 {\@qncall#1#2,}
\def\@qncall#1,{\ifunc@lled{#1}{\def\next{#1}\ifx\next\empty\else
  \w@rnwrite{Equation number \noexpand\(>>#1<<) has not been defined yet.}
  >>#1<<\fi}\else\csname @qnnum#1\endcsname\fi}

\let\eqnono=\eqno
\def\eqno(#1){\tag#1}
\def\tag#1$${\eqnono(\displayt@g#1 )$$}

\def\aligntag#1\endaligntag
  $${\gdef\tag##1\\{&(##1 )\cr}\eqalignno{#1\\}$$
  \gdef\tag##1$${\eqnono(\displayt@g##1 )$$}}

\def\eqalignno#1{\displ@y \tabskip\centering
  \halign to\displaywidth{\hfil$\displaystyle{##}$\tabskip\z@skip
    &$\displaystyle{{}##}$\hfil\tabskip\centering
    &\llap{$\displayt@gpar##$}\tabskip\z@skip\crcr
    #1\crcr}}

\def\displayt@gpar(#1){(\displayt@g#1 )}

\def\displayt@g#1 {\rm\ifunc@lled{#1}\global\advance\tagnumber by1
        {\def\next{#1}\ifx\next\empty\else\expandafter
        \xdef\csname @qnnum#1\endcsname{\t@ghead\number\tagnumber}\fi}%
  \writenew@qn{#1}\t@ghead\number\tagnumber\else
        {\edef\next{\t@ghead\number\tagnumber}%
        \expandafter\ifx\csname @qnnum#1\endcsname\next\else
        \w@rnwrite{Equation \noexpand\tag{#1} is a duplicate number.}\fi}%
  \csname @qnnum#1\endcsname\fi}

\def\ifunc@lled#1{\expandafter\ifx\csname @qnnum#1\endcsname\relax}

\let\@qnend=\end\gdef\end{\if@qnfile
\immediate\write16{Equation numbers written on []\jobname.EQN.}\fi\@qnend}

\catcode`@=12

\catcode`@=11
\newcount\r@fcount \r@fcount=0
\newcount\r@fcurr
\immediate\newwrite\reffile
\newif\ifr@ffile\r@ffilefalse
\def\w@rnwrite#1{\ifr@ffile\immediate\write\reffile{#1}\fi\message{#1}}

\def\writer@f#1>>{}
\def\referencefile{%			  Stuff to write .REF file
  \r@ffiletrue\immediate\openout\reffile=\jobname.ref%
  \def\writer@f##1>>{\ifr@ffile\immediate\write\reffile%
    {\noexpand\refis{##1} = \csname r@fnum##1\endcsname = %
     \expandafter\expandafter\expandafter\strip@t\expandafter%
     \meaning\csname r@ftext\csname r@fnum##1\endcsname\endcsname}\fi}%
  \def\strip@t##1>>{}}

\def\citeall#1{\xdef#1##1{#1{\noexpand\cite{##1}}}}
\def\cite#1{\each@rg\citer@nge{#1}}	% Variable No. of args, separated by ","

\def\each@rg#1#2{{\let\thecsname=#1\expandafter\first@rg#2,\end,}}
\def\first@rg#1,{\thecsname{#1}\apply@rg}	% each@ag is a general purpose
\def\apply@rg#1,{\ifx\end#1\let\next=\relax%	  variable no. of arg. macro.
\else,\thecsname{#1}\let\next=\apply@rg\fi\next}% args separated by commas

\def\citer@nge#1{\citedor@nge#1-\end-}	% Check for M-N range (M and N numbers)
\def\citer@ngeat#1\end-{#1}
\def\citedor@nge#1-#2-{\ifx\end#2\r@featspace#1 % Single argument
  \else\citel@@p{#1}{#2}\citer@ngeat\fi}	% M-N range of arguments
\def\citel@@p#1#2{\ifnum#1>#2{\errmessage{Reference range #1-#2\space is bad.}%
    \errhelp{If you cite a series of references by the notation M-N, then M and
    N must be integers, and N must be greater than or equal to M.}}\else%
 {\count0=#1\count1=#2\advance\count1
by1\relax\expandafter\r@fcite\the\count0,%
  \loop\advance\count0 by1\relax%	  Loop from M to N
    \ifnum\count0<\count1,\expandafter\r@fcite\the\count0,%
  \repeat}\fi}

\def\r@featspace#1#2 {\r@fcite#1#2,}	% Eat spaces at beginning or end of arg
\def\r@fcite#1,{\ifuncit@d{#1}%		  Cite individual reference
    \newr@f{#1}%
    \expandafter\gdef\csname r@ftext\number\r@fcount\endcsname%
                     {\message{Reference #1 to be supplied.}%
                      \writer@f#1>>#1 to be supplied.\par}%
 \fi%
 \csname r@fnum#1\endcsname}
\def\ifuncit@d#1{\expandafter\ifx\csname r@fnum#1\endcsname\relax}%
\def\newr@f#1{\global\advance\r@fcount by1%
    \expandafter\xdef\csname r@fnum#1\endcsname{\number\r@fcount}}

\let\r@fis=\refis			% Save old \refis, redefine
\def\refis#1#2#3\par{\ifuncit@d{#1}%      Use two params #2 #3 to strip blank
   \newr@f{#1}%
   \w@rnwrite{Reference #1=\number\r@fcount\space is not cited up to now.}\fi%
  \expandafter\gdef\csname r@ftext\csname r@fnum#1\endcsname\endcsname%
  {\writer@f#1>>#2#3\par}}

\def\ignoreuncited{%   redefine \refis if ignoring uncited references
   \def\refis##1##2##3\par{\ifuncit@d{##1}%
     \else\expandafter\gdef\csname r@ftext\csname
r@fnum##1\endcsname\endcsname%
     {\writer@f##1>>##2##3\par}\fi}}

\def\r@ferr{\endreferences\errmessage{I was expecting to see
\noexpand\endreferences before now;  I have inserted it here.}}
\let\r@ferences=\references
\def\references{\r@ferences\def\endmode{\r@ferr\par\endgroup}}

\let\endr@ferences=\endreferences
\def\endreferences{\r@fcurr=0%		  Save old \endreferences, redefine
  {\loop\ifnum\r@fcurr<\r@fcount%	  Loop over refnum and produce text
    \advance\r@fcurr by 1\relax\expandafter\r@fis\expandafter{\number\r@fcurr}%
    \csname r@ftext\number\r@fcurr\endcsname%
  \repeat}\gdef\r@ferr{}\endr@ferences}

% Save old \endpaper, redefine it to write parting message.

\let\r@fend=\endpaper\gdef\endpaper{\ifr@ffile
\immediate\write16{Cross References written on []\jobname.REF.}\fi\r@fend}

\catcode`@=12

\citeall\refto		% These macros will generate citations
\citeall\ref		%
\citeall\Ref		%

\def\Im{{\rm Im}}
\def\Re{{\rm Re}}
\def\gam{\gamma}
\def\tk{{\tilde k}}
\def\tal{{\tilde \al}}
\def\slc{SL(2,{\bf C})\ }
\def\ssc{\scriptscriptstyle}
\def\cit#1{[\cite{#1}]}

\def\cR{{\cal R}}
\def\cU{{\cal U}}
\def\cA{{\cal A}}
\def\bpart{\bar\partial}
\def\part{\partial}

\def\om{\omega}
\def\ee{{\hbox{e}}}
\def\dz{{{\rm d}^2z\over 2\pi}}
\def\dD{{\rm D}}
\def\dif{{\rm d}}

\def\al{\alpha}

\def\th{\theta}

\def\De{\Delta}
\def\de{\delta}

\def\hg{{\hat g}}
\def\gh{{\hat g}}

\def\ie{{\it i.e.,}\ }

\ignoreuncited
\rightline{iassns-hep-92-19}
\vskip -8truept
\rightline{McGill/92--26}
\vskip -8truept
\rightline{hep-th/9207117}
\font\ti=cmr10 scaled \magstep3
\title{{\ti Chiral gravity in two dimensions}}
\author{{\rm Robert C. Myers\footnote{${}^1$}{rcm@physics.mcgill.ca}}}
\affil{\rm Physics Department, McGill University
Montr\'eal, Qu\'ebec H3A 2T8, Canada}
\author{{\rm Vipul Periwal\footnote{${}^2$}{vipul@guinness.ias.edu}}}
\affil{\rm The Institute for Advanced Study
Princeton, New Jersey 08540-4920}
\abstract{It is shown that conformal matter
with $c_{\ssc L}\not=c_{\ssc R}$ can
be consistently coupled to two-dimensional `frame' gravity.
The theory is quantized, following David, and Distler
and Kawai, using the derivation of their {\it ansatz} due to
Mavromatos and  Miramontes, and D'Hoker and Kurzepa.
New super-selection rules are found by requiring SL(2,{\bf C}) invariance
of correlation functions on the plane.
There is no analogue of the $c=1$
barrier found in non-chiral non-critical strings.
A non-critical heterotic string is constructed---it
has 744 states in its
spectrum, transforming in the adjoint representation of $(E_8)^3.$
Correlation functions are calculated in this example.}

\endtopmatter
\baselineskip=14truept
%\baselineskip=20truept
\centerline{1. Introduction}
\medskip
Conformal matter in two dimensions couples to quantum gravity via
the conformal anomaly.  The coupling is characterized by the central
charge of the conformal matter.  In Euclidean space, conformal matter can be
constructed with different values of the central charge for holomorphic
and anti-holomorphic fields.  It is natural to ask how such theories,
with $c_{\ssc L}\not=c_{\ssc R},$ interact with quantum gravity.
Since these theories have
a gravitational anomaly, in addition to the conformal anomaly, there
must be degrees of freedom other than the conformal factor
that become dynamical.  This is related to an additional reason for
interest in chiral gravity---it is widely believed that {\it topological}
excitations `disorder' the world-sheet geometry when induced gravity is
coupled to matter with $c>1.$  The question then is what governs the
dynamics of such degrees of freedom.  Chiral gravity, of course, is
naturally a theory with additional {\it geometric} degrees of freedom on the
world-sheet, and its study may lead to insights into possible continuum
descriptions of physics above $c=1.$  Indeed, we shall find that
chiral gravity does involve disordering world-sheet geometry, in a precise
manner.

Recall that local gravitational anomalies can be exchanged for local
Lorentz anomalies.  A theory with a gravitational anomaly has no
generally covariant interpretation, but a theory with a Lorentz anomaly
is a theory in which the gauge invariance associated with
local frame rotations is absent.  Thus, just as the conformal
anomaly provides dynamics for the scale factor of the frame,
the Lorentz anomaly provides dynamics for local
Lorentz rotations.  In two dimensions, this is particularly
natural, for one has
$$\ee^\pm \rightarrow\ \exp\left(\rho\pm i\chi\right)\ee^\pm,
\eqno(transf)$$
under combined scale ($\rho$) and Lorentz ($\chi$) transformations.
It is the action that governs $\rho$ and $\chi$ that we shall
study in this paper.  We shall follow the
David--Distler--Kawai(DDK) approach\cit{ddk}---this has the virtue of being
applicable to arbitrary topology, and the practical advantage of
allowing the use of conformal field theory techniques.
The geometric meaning of induced chiral gravity becomes
somewhat more transparent as well.

This paper is organized as follows: in sect.~2, we set some conventions
and normalizations, and remind the reader of the calculation
of the DDK change--of--measure due
to Mavromatos and Miramontes, and D'Hoker and Kurzepa\cit{mmdk,eric}.
We show that
the analogous factor in the case of chiral gravity is the square of
that in non-chiral gravity.
In sect.~3 we use this to work out the conformal
field theory representation of chiral gravity.
In the next section we consider some global issues---it is
shown that the conformal field theory action requires careful
definition in order to preserve \slc invariance on the plane.  In
particular, the fields associated with the classical conformal or
Lorentz modes are multivalued, with super-selection rules
governing the deviation from being single-valued.
In sect.~5 we discuss operators which
are constructed purely from the fields in the gravity sector,
and examine some of the critical exponents.
In sect.~6 we give an example, the $\jmath$-string,
and compute the partition function and correlation functions in this example.
Sect.~7 contains some concluding remarks.

We note that the coupling of chiral matter to quantum gravity was
studied by Oz, Pawelczyk and Yankielowicz\cit{oz}, who worked in the
light-cone gauge introduced by Polyakov\cit{pol}.
\bigskip

\centerline{2. Review and conventions}

We will work with the Euclidean theory in this paper.  The
frame is specified locally by two 1-forms, $\{\ee^{\pm}\},$ with the
metric
$${\bf g} = {1\over 2}\big[\ee^+\otimes\ee^- +\ee^-\otimes\ee^+\big].$$
The spin connection is obtained from
$$ \dif \ee^\pm \mp i\om \ee^\pm =0.\eqno(cart)$$
The $*$ operator is defined by
$$ *\ee^\pm = \mp i \ee^\pm ,\qquad *(\ee^+\ee^-) = -2i,\qquad
*1 = {i\over 2}\ee^+\ee^-.\eqno(duoo)$$
The local area element is
${i\over 2} \ee^+\ee^-.$

Under Lorentz and conformal transformations, \(transf),
$\om$ changes as
$$\om \longrightarrow \om - *\dif\rho + \dif\chi\ .\eqno(transf2)$$
The curvature is $\cR \equiv \dif \om,$ so $\delta\cR = -\dif*\dif\rho=
*\De\rho,$ where $\De\equiv -\dif*\dif*-*\dif*\dif,$ is the
positive definite Laplacian.
$\cR$ is invariant under Lorentz
transformations, and is related to the scalar curvature $R$ by
${i\over 2}\ee^+\ee^- R = 2\cR.$

Peculiar to two dimensions is the fact that $*\om$ is a 1-form,
and from \(transf2)
$\cU\equiv\dif*\om$ behaves much as $\cR$ does, with the r\^oles of
$\rho$ and $\chi$ reversed.  Explicitly, $\delta\cU = \dif *\dif\chi =
-*\De\chi.$  Therefore $\cU$ is {\it locally} invariant under
conformal transformations. For future reference define a scalar $U$ via
${i\over 2}\ee^+\ee^- U \equiv 2\cU.$ In terms of the spin connection,
one has $U=2\nabla^\mu\om_\mu$.
The same arguments that imply that $\int\cR$
is a topological invariant imply that $\int\cU$ is also an invariant.
It is easy to see that $\int\cU=0$, since one can always
choose a frame with a divergenceless spin connection (see sect.~4).
However, we shall see that
when $\int \phi\,\cU$ occurs in an action, and $\phi$ is not
necessarily single-valued, surprising super-selection rules arise.
The `connection' $*\om$ corresponds to a reduction of the structure
group of the frame bundle to multiplication by positive real numbers,
just as $\om$ corresponds to a reduction to SO(2).

Integration over the matter degrees of freedom, using a
diffeomorphism-invariant regularization, produces an effective action
for the zweibein (see, {\it e.g.}, ref.~\cite{leu}),
$$S_{\ssc 0} =
{1\over24\pi}\left[\int *\cR{1\over \De}(c_+\cR-ic_-\cU)\right]\equiv
{c_+\over 3} S_{\ssc L} -i{c_-\over3}S_{\ssc U}\ ,
\eqno(act0)$$
where $c_\pm = (c_{\ssc R}\pm c_{\ssc L})/2.$ In background gauge,
$\ee^\pm=\exp(\rho\pm i\chi) \hat\ee^\pm,$ the dependence of $S_{\ssc 0}$
on the Weyl and Lorentz factors is purely local
$$\eqalign{
K_{\ssc L}(\rho)=S_{\ssc L}(\hat\ee,\rho,\chi)-S_{\ssc L}(\hat\ee,0,0)
&= \int \dz \left\{\part\rho\bpart\rho +{1\over4}\sqrt\gh \hat
R\rho\right\},\cr
K_{\ssc U}(\rho,\chi)=S_{\ssc U}(\hat\ee,\rho,\chi)-S_{\ssc U}(\hat\ee,0,0)
&=\int \dz \left\{-\part\rho\bpart\chi-{1\over8}\sqrt\gh \ (\hat
R\chi-\hat U\rho)\right\}.\cr}\eqno(local1)$$
An important (diffeomorphism-invariant) local counterterm can be added to the
action, namely $S_{\rm \ssc loc}={1\over 8\pi}\int \om*\om.$
The dependence on the Weyl and Lorentz factors of this term is
$$
K_{\rm\ssc loc}(\rho,\chi)=S_{\rm\ssc loc}(\hat\ee,\rho,\chi)
-S_{\rm\ssc loc}(\hat\ee,0,0)=
\int \dz \left\{ \part\rho\bpart\rho +\part\chi\bpart\chi
-{1\over4}\sqrt\gh \ (\hat U\chi-\hat R\rho)\right\}.\eqno(local2) $$
It is important to keep in mind that the effective action, \(act0),
is derived
from the anomalous conservation equation by integration.  It follows
therefore that a constant of integration can be added to
the action, if needed.  Also, this
action is {\it not} globally defined on an arbitrary surface.
Using \(local1,local2), we shall
produce a local conformal field theory expression that agrees with
eq.~\(act0) in background gauge, and then observe that the conformal
field theory is well-defined on an arbitrary surface about suitable
choices of background zweibein.  As explained in sect.~4,
additional requirements arise in the definition of the functional
integral from the requirement of conformal invariance,
that are not immediately
apparent from a consideration of the action alone.

We now briefly describe the calculation of ref.'s~\cite{mmdk,eric}.  On a
two-dimensional surface, $\Sigma,$ the
non-critical non-chiral induced gravity partition function is
$$Z \equiv \int {{\dD g}
\over {\hbox{vol.}(\hbox{Diff}_\Sigma)}}
\ \exp\left({c_m\over 3} S_{\ssc L}[g]\right)$$
where $c_m$ is the central charge of the non-chiral matter theory.
Choosing conformal gauge,  $g_{\mu\nu}=\ee^{2\rho}\hg_{\mu\nu}(m),$
and fixing diffeomorphisms \`a la Faddeev-Popov, this becomes
$$Z =\int \dif m\,\dD \rho
\ \exp\left({c_m-26\over 3} K_{\ssc L}(\rho)\right),$$
where $\dif m$ stands for an integration over the
moduli space of $\Sigma$.
The problem is now to understand the measure $\dD \rho,$
which is supposed to be the Riemannian measure induced by
$$(\de\rho,\de\rho) \equiv \int \dif^2x
\ e^{2\rho} \sqrt{\hg}\ (\de \rho)^2.$$
In ref.'s~\cite{mmdk,eric}, this problem
was tackled by comparing the problem to
finite-dimensional vector space inner products and measures related by
linear transformations.  It follows that, if $\dD_{\ssc 0}\rho$ is
the translation-invariant measure induced by
$$(\de\rho,\de\rho)_{\ssc 0} \equiv \int \dif^2x \sqrt{\hg}\ (\de \rho)^2,$$
one has $\dD\rho =
\dD_{\ssc 0}\rho\,\hbox{Det}\!\left(e^{\rho(z_1)+\rho(z_2)}
\de(z_1,z_2)\right).$  The functional determinant is singular,
but a careful heat kernel regularization\cit{mmdk} using Ward identities for
Weyl invariance\cit{eric} leads to
$$\hbox{Det} = \exp\left({1\over 3}K_{\ssc L}(\rho)\right).\eqno(det)$$

In our case, we wish to perform an integration over all
zweibein on a two-dimensional surface, $\Sigma.$ The functional
measure is defined in terms of the reparametrization invariant
distance on the space of frames given by
$$(\de\ee,\de\ee) =\int\dif^2x\,|\ee|\,{1\over2}\left(\de_{ab}\,g^{\mu\nu}
+\lambda_1\, \ee_a{}^\mu
\ee_b{}^\nu+\lambda_2\,\ee_a{}^\nu\ee_b{}^\mu\right)\,\de\ee^a{}_\mu
\,\de\ee^b{}_\nu.$$
Fixing diffeomorphisms by choosing a fiducial background frame,
$\ee^\pm=\exp(\rho\pm i\chi) \hat\ee^\pm$, leaves
$$(\de\ee,\de\ee) = \int \dif^2x
\,|\hat\ee| e^{2\rho}\,\left[(1+2\lambda_1+\lambda_2)(\de \rho)^2+
(1-\lambda_2)(\de\chi)^2\right].$$
This result is positive definite for $\lambda_2<1$ and $2\lambda_1+\lambda_2
>-1$. In the following, we will set $\lambda_1=0=\lambda_2$ since nothing
will depend on these constants. Thus variations of the conformal
and Lorentz factors can be considered independently with the separate
line elements
$$\eqalignno{
(\de\rho,\de\rho) &= \int \dif^2x \,|\hat\ee| e^{2\rho}\,(\de \rho)^2
&(dis1)\cr
(\de\chi,\de\chi) &= \int \dif^2x \,|\hat\ee| e^{2\rho}\,(\de \chi)^2
&(dis2)\cr}
$$
Changing \(dis1) to the translation invariant measure for $\rho$ is precisely
the problem discussed above, and so contributes a factor of \(det).
The measure \(dis2) on $\chi$ is already translation invariant,
and as usual, one can eliminate
the $\rho$ dependence by shifting the world-sheet metric
by a conformal factor.
This transformation yields the same factor \(det).
Hence it follows that the total determinant in this case is just the
square of \(det).
It should not come as a surprise that the Lorentz
factor does not appear in the determinant, since it does
not affect the line element.

Therefore the partition function for chiral induced gravity is
$$\eqalign{Z = &\int {{\dD \ee}\over{
{{\hbox{vol.}(\hbox{Diff}_\Sigma)}}}}
\ \exp\bigg({c_+\over 3}S_{\ssc L} - i
{c_-\over 3}S_{\ssc U} - {\xi\over 3} S_{\rm \ssc loc}\bigg)\cr
=&\int\dif m\dif n\dD_{\ssc 0}\rho\dD_{\ssc 0}\chi\
\exp\bigg({c_+-24\over 3}K_{\ssc L} - i
{c_-\over 3}K_{\ssc U} - {\xi\over 3} K_{\rm \ssc loc}\bigg)
,\cr}\eqno(final)$$
where $\dif n$ stands for a possible integration over additional Lorentz
moduli, discussed in sect.~4.
Note especially that the measure in
\(final) is not divided by the volume
of the group of Lorentz transformations---such a division would be
incorrect when $c_{\ssc L}\not=c_{\ssc R},$ since the theory is not invariant
under Lorentz transformations in this case.  This is exactly analogous
to the difference between the non-critical and critical string measures.
Also note that for the purposes of the present discussion, we have assumed
that a local counterterm has been introduced to produce a vanishing
cosmological constant on the world-sheet.
\medskip
\centerline{3. Conformal field theory}
\smallskip
We are now in a position to use conformal field theory.
Define $\phi^{\ssc
1}\equiv\rho$ and $\phi^{\ssc2}\equiv \chi.$
The functional integral in \(final) takes the form
$\int \dif m\dif n\dD_{\ssc 0}\phi^i
\ee^{-S_{\rm cft}},$ with
$$S_{\rm cft} = \int {\dif^2z\over 2\pi}\bigg[\part\phi^i\bpart\phi^j
G_{ij} -{1\over 4}\sqrt{\hg}\big(\hat R Q_j -i \hat U
P_j\big)\phi^j\bigg],\eqno(cft)$$
where
$$3G_{ij} \equiv \left(\matrix{24+\xi-c_+ &-ic_-/2\cr
-ic_-/2 & \xi\cr}\right),\quad
3Q_i \equiv\left(\matrix{-24-\xi+c_+\cr ic_-/2\cr}\right), \quad
3P_i \equiv\left(\matrix{ c_-/2\cr i\xi\cr}\right). \eqno(defq)$$
The stress tensors associated with this action are
$$\eqalign{
T &= -{1\over 2} \big[G_{ij} \part\phi^i\part\phi^j + (Q_j
+P_j)\part^2\!\phi^j\big],\cr
\bar T &= -{1\over 2} \big[G_{ij} \bpart\phi^i\bpart\phi^j + (Q_j
-P_j)\bpart^2\!\phi^j\big].\cr}\eqno(stress)$$
The fields in this theory have a non-diagonal propagator
$$\langle\phi^i(z,\bar z)\,\phi^j(w,\bar w)\rangle
=-G^{ij}\,{\rm ln}|z-w|^2\eqno(prop)$$
where $G^{ij}$ is the inverse of $G_{ij}.$
The central charge computed from
\(stress) is
$$c_{{\rm e},{\ssc R}} = 2 + 3 G^{ij}
(Q_i+P_i) (Q_j+P_j) = 26 - c_{\ssc R},\eqno(cc)$$
and similarly for $R\rightarrow L.$ Therefore in accord with expectations
\`a la DDK\cit{ddk}, the total central charge vanishes in both the holomorphic
and anti-holomorphic sectors. The parameter $\xi$ does not
appear in this expression for the central charge of the combined
conformal and Lorentz induced action.
However, $\xi$ will affect physical exponents.
Bastianelli considered a conformal field theory similar to \(cft)
for chiral bosons\cit{bas}.

The non-diagonal propagator for these fields \(prop) makes calculations
somewhat cumbersome. It is more convenient to produce a basis where
the `metric' $G_{ij}$ is diagonal, by taking linear combinations of
$\rho$ and $\chi$. (Using {\it linear} combinations produces a trivial Jacobian
in the functional measure.) Such a diagonal basis is by no means unique,
and we briefly illustrate two choices below.
We refer to these two cases as the $\rho$ theory and the $\chi$ theory,
since the bases of fields are chosen to be
$(\rho,\th\equiv\chi-ic_-\rho/2\xi)$ and $(\zeta\equiv
\rho-ic_-\chi/2(24+\xi-c_+),\chi),$
respectively.
\medskip
\item{i.} $\underline{\hbox{The}~\rho~\hbox{theory}}$
\par\noindent
In this case, we diagonalize the kinetic terms
by defining
$\th \equiv \chi -ic_-\rho/2\xi$.\footnote{$^\ddagger$}{Here, %
we assume that $\xi\not=0.$} Now
$$S_{\rm cft} = \int
\dz\bigg[{D\over 3\xi}
\big\{ \part\rho\bpart\rho +{1\over 4}\sqrt{\hg}\hat R\rho\big\}
 +{\xi\over 3}\big\{
\part\th\bpart\th-{1\over 4}\sqrt{\hg}\big(\hat U+{ic_-\over 2\xi}
\hat R\big)\th\big\}\bigg], \eqno(rhoact)$$
where $D\equiv\det 3G.$ Explicitly, one has
$${D\over\xi}=24-c_++\xi+{c_-^2\over4\xi} .
\eqno(oops)$$
If we assign $(\phi^{\ssc1},\phi^{\ssc2})=(\rho,\th)$, \(cft,stress,
prop,cc) are unchanged upon replacing \(defq) with
$$3G_{ij} \equiv \left(\matrix{D/\xi & 0\cr
0 & \xi\cr}\right),\qquad
3Q_i \equiv\left(\matrix{-D/\xi\cr ic_-/2\cr}\right), \qquad
3P_i \equiv\left(\matrix{ 0\cr i\xi\cr}\right). \eqno(newval)$$

We compute the weights of various exponentials
$$
\eqalign{\ee^{\al\rho(z)} :\ \De_\al&=-{3\over 2}{\al\xi\over D}
\left(\al-{D\over 3\xi}\right)\cr
\ee^{\tal\rho(\bar z)} :\ \De_\tal&=-{3\over 2}{\tal\xi\over D}
\left(\tal-{D\over 3\xi}\right)\cr}
\qquad
\eqalign{ \ee^{ik\th (z)} :\ \De_k &={3\over 2}{k\over \xi}
\left(k+{c_-\over 6}+{\xi\over 3}\right)\cr
\ee^{i\tk\th(\bar z)}:\ \De_\tk &={3\over 2}{\tk\over \xi}
\left(\tk+{c_-\over 6}-{\xi\over 3}\right)\cr}
\eqno(dress11)$$
Above, we have explicitly introduced separate momenta for the holomorphic
and anti-holomorphic exponentials because the discussion in the next section
suggests that we should be considering operators which introduce
cuts.\footnote{$^\dagger$}{In sect. 4, we will find %
$\tal=\al$, so that the formul\ae\  presented here are slightly more general %
than needed.} Even if $\tk=k$, $\De_k-\De_\tk=k$.
This spin for an operator $\exp[ik(\th_{\ssc R}+\th_{\ssc L})]$
results from the coupling of $\theta$ to $\hat U$.

Given a matter operator of weight $(\De_{\ssc L},\De_{\ssc R}),$ one may
expect that it acquires exponential dressings by $\rho$ and $\th$
to make it a (1,1) operator, as in non-chiral gravity\cit{ddk}.
For a given $k$, $\al$ is determined to be
$$\al
={1\over 2} \left[{D\over3\xi}\pm\sqrt{{D^2\over9\xi^2}-{8D\over3\xi}
\left(1-\Delta_{\ssc R}-\De_k\right)}\,\right],\eqno(dress1)$$
or alternatively for fixed $\al$, $k$ is
$$k
=-{1\over 6} \left[{c_-\over2}+\xi\pm\sqrt{
\left({c_-\over2}+\xi\right)^2+24\xi
\left(1-\Delta_{\ssc R}-\De_\al\right)}\,\right].\eqno(dress2)$$
Therefore one finds a one parameter family of dressings in the holomorphic
sector. Similar formul\ae\  relate $\tal$ and $\tk$ in the anti-holomorphic
sector.
While two branches have been indicated in eq.'s \(dress1,dress2)\ for possible
gravitational dressings, the negative sign yields the correct results
in the limit $\xi\rightarrow\infty$.
We will show that the latter corresponds to
the classical limit of chiral gravity in sect.~7.
The exponential dressings we have considered here are the simplest
possible dressings.  Because of the spin inherent in the $\th$ exponentials,
we expect that it is possible to find other primary fields constructed
from $\rho$ and $\th$, which can dress matter operators to
produce weight (1,1) operators (see sect.~5).

\item{ii.} $\underline{\hbox{The}~\chi~\hbox{theory}}$
\par\noindent
An alternate diagonalization comes from the choice
$(\phi^{\ssc1},\phi^{\ssc2})=(\zeta\equiv
\rho-ic_-\chi/2X,\ \chi)$ where $X\equiv 24+\xi-c_+.$
In this case, one replaces \(defq) with
$$3G_{ij} \equiv \left(\matrix{X& 0\cr
0 & D/X\cr}\right),\qquad
3Q_i \equiv\left(\matrix{-X\cr 0\cr}\right), \qquad
3P_i \equiv\left(\matrix{ c_-/2\cr iD/X\cr}\right), \eqno(defq2)$$
in \(cft,stress,prop,cc).
$\chi$ couples only to $\hat U,$ and the coupling of
$\zeta$ to $\hat U$ vanishes as $c_-\rightarrow0.$ Also,
in contrast to the $\rho$ theory, the limit $\xi\rightarrow0$ is
nonsingular in this case.

As in the previous case, one can calculate gravitational dressings
for matter fields. Here, we will only note that in general
the weight of an operator is given by
$${\left[\ee^{k_i\phi^i}\right]_{\ssc R} = -{1\over 2}
\left\{G^{ij}k_i(k_j+Q_j+P_j)\right\}},$$
while replacing $P_j$ by $-P_j$ above yields the weight of an
exponential of anti-holomorphic fields.
\def\lor#1#2{#1\leftrightarrow #2}

Finally, we comment on a duality between these two diagonalized theories.
The following substitutions in \(defq2) take the $\chi$ theory to the $\rho$
theory:
$$\chi\rightarrow{-i\rho},\quad\zeta\rightarrow{i\th},
\quad\lor X{-\xi},\quad\lor{\hat U}{i\hat R}.$$
Further considering the weights of exponential operators in the two theories,
one finds
$$ \left[\ee^{\al\rho+ik\th}\right]_R=\left[\ee^{k\zeta+i\al\chi}\right]_R
\qquad\hbox{and}\qquad
\left[\ee^{\tal\rho+i\tk\th}\right]_L=\left[\ee^{-\tk\zeta-i\tal\chi}\right]_L
\ ,$$
with the above substitution $\lor X{-\xi}.$ Thus
the $\chi$ and $\rho$ theories are isomorphic for appropriate
values of $X$ and $\xi.$

\medskip
\centerline{4. Global issues}

Thus far we have evaluated local quantities in the conformal field
theory, and have therefore had no need to consider the definition of
the terms in the action on a non-trivial manifold.  {\it A priori}
it is not clear that expressions such as
$\int \om *\om$ can be defined in a invariant manner over the
entire surface considered.  We will now demonstrate that the
conformal field theory action is well-defined on
any orientable surface.

The problem is as follows: On a higher genus surface, the action, \(cft), is
evaluated by a sum of integrals on a set of overlapping coordinate patches.
In general on an overlap between adjacent patches, the background
zweibeins $\hat\ee^\pm$ are related by {\it both} coordinate and Lorentz
transformations. Now in order to be well-defined, the action should
be unchanged if the boundaries between the integrals are shifted. This
is clearly the case for the kinetic terms and the couplings to $\hat R$,
which are Lorentz invariant and covariant under coordinate changes.
Recall though that under a Lorentz transformation, $\de
\hat U=2\,*\dif *\dif\hat\chi$. Thus moving boundaries would appear to
change the contributions of the couplings to $\hat U$. A simple
resolution of this problem is easily found since we are only interested in
complex coordinates for conformal field theory.
Gauss showed that locally there exist coordinates such that
the metric is conformally flat,
$${\bf\hat g} =\hbox{$1\over 2$} {\ee^{2\hat\rho}}\left(
\dif z\otimes\dif\bar z
+\dif \bar z\otimes\dif z\right) .$$
This implies that locally there exist zweibein such that
$\hat U=0,$ since we can choose
$$\ee^+ \equiv \ee^{\hat\rho} \dif z
\quad{\rm and}\quad
\ee^- \equiv \ee^{\hat\rho} \dif\bar z\,.\eqno(guas)$$
In general, one can allow for additional holomorphic Lorentz
transformations between patches. In either case, $\hat U$ is invariant
in the overlaps leading to a well-defined action.
Henceforth, we make the choice of a Gaussian background
zweibein \(guas) with $\hat U=0$.

\def\tchi{\tilde\chi}

It remains to determine the moduli associated with the integration
over `all zweibein'.  Since phases cancel when zweibein are tensored to
produce the metric, there are global phases that must be integrated over,
above and beyond the moduli associated with the integration over
conformal equivalence classes of metrics. In the {classical} geometry,
one can realize these phases by shifting the spin connection,
$\om\rightarrow\om+\sum_{i=1}^{2g}\lambda_i\,\beta^i$, where $\beta^i$ is a
basis for the harmonic differentials on the genus $g$ surface. Since
$\beta^i$ are closed and divergenceless, both $\cR$ and $\cU$ are
uneffected by this shift. Given any closed contour around a particular
nontrivial cycle though, one acquires an additional phase:
$\int_a\om\rightarrow\int_a\om+\lambda_a$.
These global phases are most conveniently incorporated into the Lorentz field
$\chi$. To be precise, one lets
$$\dif\chi=\dif\tchi+\sum_{i=1}^{2g}\lambda_i\,\beta^i\ .$$
where $\tchi$ is a (single-valued) function on the surface.
In contrast since the $\beta^i$
are not exact forms, $\chi$ must now be multivalued on the surface
(or alternatively, $\chi$ contains discontinuities).
The measure for the Lorentz moduli,
$\dif n=\prod\dif\lambda_i$, would then be
included as a part of the functional measure $D\chi$.
These global phases are associated with the nontrivial cycles on the
higher genus surfaces. Additional nontrivial cycles occur  in correlation
functions surrounding the operator insertions. The above discussion
then suggests that we allow $\chi$ to have discontinuities around such cycles.
Such cuts would be produced by dressing the matter operators with
exponentials of the form: $\exp[\al\rho+ik\chi_{\ssc R}+i\tk\chi_{\ssc L}]$.
Below, we will see that these expectations motivated by the classical
geometry must be slightly modified in the quantum theory.

Usually momentum (non)conservation super-selection rules for correlation
functions are derived in the path integral framework. One would consider
integrating the constant mode of the fields $\phi^i$. These modes
vanish in the kinetic terms, and in the $\hat U$ interactions as well since
$\int\hat\cU=0$. The coupling to the background curvature gives a
contribution proportional to the Euler constant of the surface. Thus
in a correlation function with a number of exponential insertions
$\exp[k^{\ssc (a)}_i\phi^i]$, integrating the constant mode
(over an appropriate contour) yields
a delta function requiring $\sum_a k^{\ssc(a)}_i=-Q_i(1-g).$
A slightly more general calculation allowing for multivalued fields
produced by operator insertions $\exp[k^{\ssc (a)}_i\phi_{\ssc R}^i
+\tk^{\ssc (a)}_i\phi_{\ssc L}^i]$, would appear to
yield $\sum_a (k^{\ssc(a)}_i
+\tk^{\ssc (a)}_i)=-2Q_i(1-g)$ and $\sum_a (k^{\ssc(a)}_i
-\tk^{\ssc (a)}_i)=0.$
In fact, we show below that these naive results are incorrect.

The correct momentum conservation rules can be deduced by a consideration of
\slc invariance on the plane. Consider a correlation function
of exponential operators. For the purposes of an explicit example, we
assume that the matter theory is a set of chiral bosons $X^i$ with no
background charges. The holomorphic part of the operators takes the
form, $\exp[k^{\ssc (a)}_i\phi^i+i\gam^{\ssc (a)}_i X^i]$, where
demanding that they have weight 1 requires
$$2=\gam^{\ssc (a)}_i\gam^{\ssc (a)}_i-G^{ij}k^{\ssc (a)}_i(k^{\ssc (a)}_j
+Q_j+P_j)\ \ .\eqno(wate)$$
Correlation functions may be written as $\cA\equiv\int \cA_{\ssc
R}\cA_{\ssc L},$ with
the holomorphic part of the correlation function measure taking the form
$$\cA_{\ssc R}=\prod_a\dif z^{\ssc (a)}\prod_{a<b}\left(z^{\ssc (a)}-
z^{\ssc (b)}\right)^{\De_{ab}}\ ,
\quad{\rm where}\quad
\De_{ab}=\gam^{\ssc (a)}_i\gam^{\ssc (b)}_i-G^{ij}k^{\ssc (a)}_i
k^{\ssc (b)}_j\ .$$
Now $\cA_{\ssc R}$ should be invariant\footnote{${}^*$}{In general, of %
course, $\cA_{\ssc R(L)}$ need only be invariant up to complex %
conjugate phases.} under an \slc
transformation: $z^{\ssc (a)}\rightarrow(a z^{\ssc (a)}+b)/(c z^{\ssc (a)}
+d)$ where $ad-bc=1$. This transformation takes $\dif z^{\ssc (a)}\rightarrow
\dif z^{\ssc (a)}/(c z^{\ssc (a)}+d)^2$, $\left(z^{\ssc (a)}-
z^{\ssc (b)}\right)\rightarrow\left(z^{\ssc (a)}-z^{\ssc (b)}\right)/[
(c z^{\ssc (a)}+d)(c z^{\ssc (b)}+d)]$, and hence
$$\cA_{\ssc R}\rightarrow\cA_{\ssc R}\prod_a(c z^{\ssc (a)}+d)^{-2}
\prod_{e<f}\left[(c z^{\ssc (e)}+d)(c z^{\ssc (f)}
+d)\right]^{-\De_{ef}}\ \ .$$
Therefore invariance of $\cA_{\ssc R}$ requires for each $a$
$$2=-\sum_{b\not=a}\De_{ab}
=-\gam^{\ssc (a)}_i\sum_{b\not=a}\gam^{\ssc (b)}_i+
G^{ij}k^{\ssc (a)}_i\sum_{b\not=a}k^{\ssc (b)}_j\ \ .\eqno(conq)$$
Momentum conservation for the matter fields requires $\sum_{b}
\gam^{\ssc (b)}_i=0$, while similarly in the gravity theory
one expects to find $\sum_{b}k^{\ssc (b)}_i=-\Lambda_i$ for some
constants $\Lambda_i$. Using these conservation equations, \(conq)
becomes
$$2=\gam^{\ssc (a)}_i\gam^{\ssc (a)}_i-G^{ij}k^{\ssc (a)}_i(k^{\ssc (a)}_j
+\Lambda_j)\ \ ,$$
which is precisely the weight 1 condition \(wate) when $\Lambda_i=Q_i+P_i$.
Hence we have deduced the superselection rule
$$\sum_{a}k^{\ssc (a)}_i=-Q_i-P_i\eqno(hollow)$$
in the holomorphic sector. The same calculation for anti-holomorphic
exponentials, $\exp[\tk^{\ssc (a)}_i\phi^i
+i\tilde\gam^{\ssc (a)}_i X^i]$, yields
$$\sum_{a}\tk^{\ssc (a)}_i=-Q_i+P_i\ \ .\eqno(ahollow)$$
These results are in contradiction with those produced by a naive path
integral approach since here we have $\sum_{a}(k^{\ssc (a)}_i-\tk^{\ssc (a)}_i)
=-2P_i$, which is nonvanishing in general. A more careful
treatment of the $\hat U$ coupling in the presence of multivalued
fields may reproduce \(hollow,ahollow) from a path integral calculation.

Now if one does not allow any cuts to occur in the gravity sector fields
(\ie $k^{\ssc (a)}_j=\tk^{\ssc (a)}_j$), all correlation functions in the
plane will vanish since \(hollow) and \(ahollow) cannot then be
satisfied simultaneously.
One might only insist $\rho$ have no cuts, but allow $\chi$
to be multivalued following the considerations of the classical geometry
given above. All planar correlation functions vanish in this case as well, as
can be seen from the discussion of the original basis of fields in sect.~3
where $\rho$ couples to $\hat U$ with $P_{\ssc 1}=c_-/6.$

These problems arise because of the $\hat U$ interactions in \(cft).
Thus we are lead to the natural suggestion that one should allow
for discontinuities
in the linear combination of fields that couples to $\hat U$,
namely $\th=\chi-ic_-\rho/2\xi$. We will also insist $\rho$ be
single-valued, as in the classical geometry. Thus the $\rho$ theory
in sect.~3
provides the natural basis in which to study correlation functions.
The exponential dressings determined from \(dress11,dress1,dress2)
should have $\tal=\al$, but this still leaves one free parameter,
namely the phase or cut introduced in $\th$.
Explicitly, the super-selection rules are
$$\eqalign{
\sum_{a} \al^{\ssc (a)} &= {D\over3\xi} (1-g), \cr
\sum_{a} k^{\ssc (a)} &= \left[-{c_-\over6}-{{\xi}\over 3}
\right] (1-g), \cr
\sum_{\ssc a} \tk^{\ssc (a)} &= \left[-{c_-\over6}+{{\xi}\over 3}\right]
(1-g). \cr}\eqno(ssrho)$$
These rules are extended to arbitrary genus surfaces,
as would result by standard factorization
arguments.  Such an extension is easily verified to be correct for
$g=1.$  Finally note that the original discussion of the Lorentz moduli arising
on such higher genus surfaces should be modified by replacing the
field $\chi$ by $\th$.

On a Euclidean world-sheet, it is natural to identify $\chi$ with
$\chi+2\pi$ ({\it i.e.}, a constant $2\pi$ rotation leaves the
zweibein invariant). Implementing this identification
results in various quantization conditions. For instance,
$c_- \in 6{\bf Z},$ so that $\ee^{-S_{\rm cft}}$ only acquires
a phase of $\exp(2\pi\,i\,n)$ when $\chi$ is shifted by $2\pi$,
due to the curvature interaction.
We will ignore any such
identification though, because it would be incorrect
if one regards this theory as the analytic continuation of
a Minkowskian world-sheet theory, where the Lorentz group is
non-compact.

Global Lorentz and diffeomorphism anomalies are distinct.  Global
diffeomorphism invariance translates into the modular invariance
of the partition function defining the theory, while global Lorentz
transformations are not an issue for the non-compact form of the
Lorentz group. In sect.~6 we shall see that
rather intriguing theories are selected by this criterion
of modular invariance.

\medskip
\centerline{5. More operators}
\smallskip
The classical theory had zweibein and a metric,
so one might be interested in
operators in the quantum theory with similar transformation properties.
For instance, the zweibein transform as fields of weight (1,0) or (0,1).
{}From \(dress11), we see that a (1,0) operator of the form,
$\exp[\al\rho+ik\th_{\ssc R}+i\tk\th_{\ssc L}]$, has
$$
\eqalign{
k&=-{1\over 6} \left[{c_-\over2}+\xi\pm\sqrt{\left({c_-\over2}+\xi\right)^2
-24\xi\De_\al}\,\right]\cr
\tk&=-{1\over 6} \left[{c_-\over2}-\xi\pm\sqrt{\left({c_-\over2}-\xi\right)^2
-24\xi\left(\De_\al-1\right)}\,\right].\cr} %
$$
Such operators will prove useful as dressing fields in the next section.
Due to the spin inherent to $\th$ exponentials a solution
introducing no discontinuities is possible with $k=\tk=-1$.
Similarly a weight (1,1) operator is produced if in the expression for $k$
above, $\De_\al$ is replaced by $(\De_\al-1)$. These operators
correspond to introducing a puncture in the surface, which is also
the origin of a cut in $\th$. In this case, producing a local operator (\ie
no cuts) requires that $k=\tk=0$, and
$$\al
={1\over 6} \left[{D\over\xi}\pm\sqrt{{D\over\xi}\left({D\over\xi}-24\right)
}\,\right].\eqno(comic)$$
Agreement with the classical limit, $\xi\rightarrow\infty$, requires
choosing the negative branch. This local operator provides a
cosmological constant operator, which could be added as a new perturbative
interaction to \(cft). It may be of
interest to investigate the deformed theory
{\it \`a la} Goulian and Li\cit{mark}.

This cosmological constant operator is the quantum analog of an operator
that exists in the classical geometry. One might ask if there are other
classical operators, which may have quantum analogs in this way. The answer
is an affirmative, since we need only insist that these operators be
diffeomorphism invariant, but they need be neither Weyl nor Lorentz
invariant. With only one derivative, one might consider: $\dif \ee^\pm$,
$\dif\!*\!\ee^\pm$, $\om \ee^\pm$, and $\om\!*\!\ee^\pm$. In the classical
geometry using
\(cart,duoo), one shows that only two of these operators are distinct:
$\dif \ee^+=\bpart(\ee^{\rho+i\chi})\dif\bar z\dif z$ and
$\dif \ee^-=\part(\ee^{\rho-i\chi})\dif z\dif\bar z$. Unfortunately, these
are total derivatives locally,
and hence will not provide interesting interactions.
One can confirm that in the quantum gravity theory there are two
primary (1,1) operators in the form of an exponential and a single
holomorphic or anti-holomorphic derivative, that these operators are indeed
total derivatives (except for special values of $\xi$), and
that they do in fact coincide with the above expressions in the classical
limit, $\xi\rightarrow\infty$.

\def\derv{\nabla_{\!z}}
\def\bderv{{\nabla}_{\!\bar z}}
One might consider classical operators of the general form
$$[\derv\, ;n]\,[\bderv\, ;m]\,\exp(\al\rho+ik\chi)\,\dif z\dif\bar z
\eqno(doood)$$
where $[\derv\, ;n]$ (resp. $[\bderv\, ;m]$) is a sum of arbitrary terms each
containing a total of $n$ (resp. $m$ anti-) holomorphic covariant
derivatives acting on $\rho$ and/or $\chi$.
Following ref.~\cite{fried}, one constructs operators which are
invariant under conformal reparametrizations:
In a local complex coordinate patch, the nonvanishing
components of the zweibein and inverse-metric are $\ee^+{}_z
=\ee^{\rho+i\chi}$,
$\ee^-{}_{\bar z}=\ee^{\rho-i\chi}$ and $g^{z\bar z}=2\ee^{-2\rho}$.
Since $[\derv\, ;n]$ acts as a tensor of weight (0,$n$) under analytic
reparametrizations, one can form an invariant scalar,
$(\ee^-{}_{\bar z}\,g^{z\bar z})^n[\derv\, ;n]\sim \ee^{-n\rho-in\chi}
[\derv\, ;n]$. Similarly, $(\ee^+{}_{z}\,g^{z\bar z})^m[\bderv\, ;m]\sim
\ee^{-m\rho+im\chi}[\bderv\, ;m]$
and $\ee^+\ee^-\sim \ee^{2\rho}\dif z\dif\bar z$
are invariant. Combining these formul\ae, one constructs
an invariant operator in \(doood) by setting
$$k=m-n\qquad\qquad{\rm and}\qquad\qquad\al=2-m-n\ .\eqno(restr)$$
As a simple example, $n=2$,
$m=0$ yields $k=-2$, $\al=0$, and $[\derv\, ;2]$ is a sum of five possible
terms, but
two linear combinations yield total derivatives. Hence there are three
new nontrivial interactions, which might have analogs in the quantum
theory. Working with the basis, $(\phi^{\ssc1},\phi^{\ssc2})=(\rho,\th)$,
in the conformal field theory,
one finds a single (1,1) operator of the form
$$ \left[\eta_{ij}\part\phi^i\part\phi^j+\psi_i\part^2\!\phi^i\right]
\exp\left(q_i\phi^i\right)$$
where $q_{\ssc2}=ik=-2i$ and
$$q_{\ssc1}=\al
={1\over 2} \left[{D\over3\xi}-\sqrt{{D^2\over9\xi^2}+{8D\over3\xi}
\left(\De_k+1\right)}\,\right].$$
Here only the solution for $\al$, with the correct classical limit as
$\xi\rightarrow\infty$, is displayed.
Writing the `polarization' tensor as
$\eta=\hat\eta+{1\over2}(q\otimes\psi+\psi\otimes q)$,
one also requires that
$$\hat\eta_{ij}G^{jk}(Q_k+P_k+q_k)={1\over6}G^{jk}\hat\eta_{jk}q_i\qquad
{\rm and}\qquad G^{ij}\hat\eta_{ij}+3G^{ij}\psi_i(Q_j+P_j+q_j)=0\ .$$
Generically, the solution of these equations is unique up to the homogeneous
solution
$\psi_{\ssc 0}$ of the latter equation (\ie set $\hat\eta=0$), which
corresponds to a total derivative operator. The spin exponent
$k=-2$ is not renormalized in the quantum operator. This property
holds in general
for such operators. Thus we have found a quantum analog for one
of the three classical expressions. Presumably more quantum interactions
could be constructed if one considers a general ansatz which
allows for mixing with the background connection and curvature\cit{dp},
or alternatively with the ghost current\cit{bank}.

Apparently these new operators are examples of discrete states in chiral
gravity. The determination of the complete spectrum is beyond the scope
of the present paper. It should be noted that the spectrum appears to depend
on $\xi$ in a complicated manner. Further the above discussion focussed
on local operators, which might appear as interactions in the action.
When one allows nontrivial phases in $\th$, a continuum of new operators
is produced for each one of these interactions. It may be of interest
to investigate the analog of the Witten ground ring\cit{good} in these
theories.

We conclude this section by briefly addressing the question of the
critical `dimension' for chiral gravity. One might begin by demanding
the reality of the string susceptibility as in
\cit{ddk}. Define the area operator,
$\int\!\dif^2x\,\sqrt{\hat{g}}\,\ee^{\al\rho}$, with $\alpha$
given by \(comic). This definition is chosen to involve only
a local dressing operator,
and to make no explicit reference to the Lorentz field, which should
be irrelevant to defining the area, in analogy to the classical geometry.
Then one considers the fixed area partition function
$$Z(A)=\left\langle\,\delta\left(\int\!\dif^2x\,\sqrt{\hat{g}}\,
\ee^{\al\rho} -A\right)\,\right\rangle=0\ \ .$$
Here a vanishing result occurs, except for genus one surfaces, because the
super-selection rules for $\th$ in \(ssrho) are not satisfied. (For $g=1$,
one finds $Z(A)\propto A^{\Gamma-3}$ with $\Gamma=2$, exactly as in
nonchiral gravity\cit{ddk}.) To properly fix the {\it two} $\th$ zero mode
integrals, one can consider inserting punctures with dressings which
absorb the appropriate $\th$-momenta. If one introduces a single puncture,
there is a unique dressing which yields a nonzero
result for $g=0,1$. $P_{\ssc 0}
=\int\!\dif^2x\sqrt{\hat{g}}\exp[\beta\rho+ik
\th_{\ssc R}+i\tilde k\th_{\ssc L}]$ with $\beta=\al$ precisely
as in the area operator, and $k=(g-1)(c_-/6+\xi/3)$
and $\tilde k=(g-1)(c_-/6-\xi/3)$. Now one has
$$Z'(A)=\left\langle\,P_{\ssc 0}
\ \delta\left(\int\!\dif^2x\,\sqrt{\hat{g}}\,\ee^{\al\rho}
-A\right)\,\right\rangle\propto A^{\Gamma-2}$$
where
$$\Gamma=2+{g-1\over12}\left[{D\over\xi}
+\sqrt{{D\over\xi}\left({D\over\xi}-24
\right)}\,\right]\eqno(ggaamm)$$
for $g=0,1.$
This is precisely analogous to the result for nonchiral gravity\cit{ddk}
with the replacement $(25-c_m)\rightarrow D/\xi$.
If we make the assumption that $D/\xi$ is positive so that the $\rho$-action
\(rhoact) has a positive coefficient, then from \(oops)
requiring that $\Gamma$ be real imposes the restriction
$${{D\over\xi}-24}\ \ =\ \ \xi+{c_-^2\over4\xi}-c_+>0\ \ .\eqno(jacket)$$
Now for any fixed values of $c_{\pm}$, there will exist values of $\xi$
for which this inequality is satisfied. Therefore chiral gravity has no
barriers for the allowed values of central charges in the matter sector.

Of course, this conclusion relies on results for only $g=0,1,$ since
$Z'(A)$ always vanishes for $g>1$. Introducing two punctures yields nontrivial
results for higher genera as well. In this case, one finds
that $Z''(A)\propto A^{\widetilde\Gamma-1}$ where $\widetilde\Gamma
=\Gamma+(\al_++\al_--2\al)/\al$. Here, $\Gamma$ is given by
\(ggaamm), while
$\al_\pm={1\over 6}\left[{D/\xi}+\sqrt{{D}Y_\pm/\xi}\right]$ with
$$Y_\pm=\left[2g^2-1\pm2g\sqrt{g^2-1}\,\right]\xi+
        \left[2g^2-1\mp2g\sqrt{g^2-1}\,\right]{c_-^2\over4\xi}-c_+$$
for $g\ge1$. Thus one finds that the genus dependence of the string
susceptibility is far more complicated than the simple linear dependence
found in nonchiral gravity\cit{ddk}. Further requiring real exponents
by imposing $Y_\pm>0$, only produces constraints which are less restrictive
than \(jacket). Finally note that we have ignored ghost zero
modes in this entire discussion, but that a more rigorous account can be
given completely equivalent to that appearing in ref.~\cite{ddk}.

\medskip
\centerline{6. The $\jmath$--string}
\smallskip
We now construct a simple example of a heterotic non-critical string.
Consider the holomorphic conformal field theory associated with the
$E_{\ssc 8}\times E_{\ssc 8}\times E_{\ssc 8}$ root lattice.  One may
view this as described by 24 chiral bosons, or as 48 chiral fermions
with appropriate sums over spin structures. This choice has $c_\pm=12$,
which simplifies calculations of correlation functions.
Many choices of 24--dimensional chiral lattices are possible (see, {\it e.g.},
ref.~\cite{shell}), but we choose $(E_8)^3$ for concreteness, and since
it gives the amusing result that the
partition function is the $\jmath$ invariant\cit{jps} on the
torus.

To calculate the partition function, at genus 1, one must
account for several contributions\cit{dp}. First, the modular invariant
measure is $\dif^2\tau/(\Im\tau)^2$. The contribution coming from
the Faddeev-Popov Jacobian, for fixing diffeomorphisms,
is: $\Im\tau|\eta(q)|^4$, where
$q\equiv \exp(2\pi i\tau),$ and $\eta(q)$
is the Dedekind $\eta$-function\cit{whit}.
The integration over the matter fields yields
$$\jmath(\tau) \equiv \left({\Theta_{E_8}(q)\over\eta(q)^8}\right)^3=
\left({\Theta_2(q)^2+\Theta_3(q)^2+\Theta_4(q)^2\over\eta(q)^8}\right)^3$$
where $\Theta_i(q)=\Theta_i(z=0|\tau)$ are Jacobi theta functions\cit{whit}.
The $\rho$ contribution is the same as that for a free scalar field:
$(\Im\tau)^{-1/2}|\eta(q)|^{-2}$.\footnote{$^\dagger$}{We ignore the %
unconstrained integrals over the $\rho$ and $\theta$ zero modes in the %
present discussion.}

The $\th$ contribution is more interesting because of the Lorentz
moduli. Covering the torus with a fixed coordinate patch,
$0\le\sigma^{\ssc1},\sigma^{\ssc2}\le1$, the world-sheet metric becomes
$\dif s^2=|\dif\sigma^{\ssc1}+\tau\,\dif\sigma^{\ssc2}|^2$.
We introduce arbitrary phases in $\th$
around the $\sigma^i$ cycles by setting
$$\th(\sigma^{\ssc1},\sigma^{\ssc2})
=\tilde\th(\sigma^{\ssc1},\sigma^{\ssc2})+2\pi\lambda_i\sigma^i.$$
Here $\tilde\th$ is a function on the torus, and the second term
incorporates the multivalued contribution of the Lorentz moduli.
The contribution to the partition function is then found to be
$$Z_\th={1\over(\Im\tau)^{1/2}|\eta(q)|^2}\int_{-\infty}^{\infty}
\dif\lambda_{\ssc1}\dif\lambda_{\ssc2}
\ \exp\left(-{\pi\xi|\lambda_{\ssc2}-\lambda_{\ssc1} %
\tau|^2\over6\,\Im\tau}\right)$$
where we have included the phase integral, which one may note is
invariant under modular transformations. The integral takes
a particularly simple form upon shifting $\hat\lambda_{\ssc2}=
\lambda_{\ssc2}-\lambda_{\ssc1}\Re\tau$ %
$$\eqalign{
Z_\th=&{1\over(\Im\tau)^{1/2}|\eta(q)|^2}\int_{-\infty}^{\infty}
\dif\lambda_{\ssc1}\dif\hat\lambda_{\ssc2}
\ \exp\left(-{\pi\xi\over6}\left\{{\hat\lambda_{\ssc2}^2\over
\Im\tau}+\Im\tau\lambda_{\ssc1}^2\right\}\right)\cr
=&{1\over(\Im\tau)^{1/2}|\eta(q)|^2}\ {6\over\xi}\cr}
\eqno(lorry)$$
Therefore after the phase integral is performed, the final $\th$ contribution
is also simply that of a free scalar field (up to an arbitrary normalization).

Combining all of these contributions yields
$$Z_{\ssc\rm torus} = \int_{\cal F} {\dif^2\tau\over(\Im\tau)^2}
\ \jmath(\tau)\eqno(torepart)$$
where $\cal F$ is a fundamental domain for
the action of the modular group on the Poincar\'e upper-half-plane.
To understand the spectrum
of the theory, we investigate the region of
small $q,$ equivalently large $\Im\tau,$ where
$$\jmath(\tau)\sim {1\over q} + 744 +\dots \qquad
.\eqno(744)$$
The integration over $\Re\tau$ projects out any term in eq.~\(744) with
a non-zero power of $q,$ including the tachyon.
The only term left is the constant $744,$ indicating that there are
$744$ states in this string theory, $720$ corresponding to
winding number states in the $24$--dimensional lattice, and $24$ coming
from the maximal torus of the group, the so-called oscillator states.
If one counted the states before performing the phase integral
in \(lorry),
each of the 744 states corresponds to a continuum with arbitrary phases
around the $\sigma^{\ssc1}$ cycle.

We now consider some sample correlation functions in this theory.
{}From the partition function, it appears that the physical states
correspond to (0,1) operators in the matter sector. Thus we
construct exponential (1,0) dressings, as discussed in sect.~5.
First though in the present theory with $c_\pm=12$
in the ($\rho,\th$) basis, $G_{ij}={\rm diag}\left((x+2/x)^2,x^2
\right)$ where $x=\sqrt{\xi/3}$. Therefore it is extremely convenient
to rescale the fields to a new basis ($\hat\rho=(x+2/x)\rho,\,\hat\th=x\th$)
for which
$$G_{ij} =\de_{ij},\qquad
Q_i \equiv\left(\matrix{-(x+2/x)\cr 2i/x\cr}\right), \qquad
P_i \equiv\left(\matrix{ 0\cr ix\cr}\right). $$
Now there are four possible (0,1) operators of the form
$\exp[\al\hat\rho+ik\hat\th_{\ssc R}+i\tk\hat\th_{\ssc L}]$, but we restrict
our attention to the two with good semiclassical limits.
For fixed $k$, they correspond to
\item{i.}$\tk=k+x,\ \al=-k$
\item{ii.}$\tk=-2/x-k,\ \al=-k\ \ .$
\par\noindent
The first introduces a fixed phase in the $\th$ field with
$k-\tk=-x$, while the phase of the second dressing varies continuously
with $k$, $k-\tk=2/x+2k$. The momentum super-selection rules for correlation
functions are: $\sum\al^{\ssc (a)}=x+2/x$, $\sum k^{\ssc (a)}=-x-2/x$
and $\sum \tk^{\ssc (a)}=x-2/x$.

One can begin by considering three-point amplitudes of some given
set of matter operators. For a fixed set, there would be eight
possible amplitudes corresponding to all of the different combinations
of dressings. Some of these amplitudes vanish since they combine
dressings which are incompatible with the momentum conservation rules.
The amplitudes which survive are those which involve either
two (i) and one (ii) dressings, or one (i) and two (ii).
In those cases where the amplitude is non-vanishing, one
finds that as expected the results are \slc invariant, but
also {\it independent} of
how the various dressings are combined with the matter operators.
There are two classes of nonvanishing amplitudes: those with three
winding number states, $\exp(i\gamma^{\ssc(a)}\!\cdot\! X_{\ssc R})$,
which yield
simply $\delta(\gamma^{\ssc(1)}+\gamma^{\ssc(2)}+\gamma^{\ssc(3)})$,
and those with two winding number states and one oscillator state,
$i\beta\!\cdot\!\part X_{\ssc R}$, which yield
$\beta\cdot\gamma^{\ssc(1)}\ \delta(\gamma^{\ssc(1)}
+\gamma^{\ssc(2)})$.\footnote{$^\dagger$}{These results use %
Kronecker $\delta$-functions, since the winding number vectors are discrete.}

The four-point amplitudes produce more interesting results, as we
illustrate here. Consider the following correlation function
$$\eqalign{\cA=\int\dif^2z^{\ssc(3)}\ \Big\langle
c\bar c\, V_{\rm i}\, \ee^{i\gamma^{\ssc(1)}\cdot X_{\ssc R}}(z^{\ssc(1)})
\ c&\bar c\, V_{\rm i}\, \ee^{i\gamma^{\ssc(2)}\cdot X_{\ssc R}}(z^{\ssc(2)})
\cr\times  &V_{\rm ii}\, \ee^{i\gamma^{\ssc(3)}\cdot X_{\ssc R}}(z^{\ssc(3)})
\ c\bar c\, V_{\rm i}\, \ee^{i\gamma^{\ssc(4)}\cdot
X_{\ssc R}}(z^{\ssc(4)})\Big\rangle\ ,\cr}$$
where $c\bar c$ are ghost dressings for the fixed operators, $V_{\rm i(ii)}$
are gravitational dressings of type i (ii) given above, and
$\exp[i\gamma^{\ssc(a)}\cdot X_{\ssc R}]$ are winding state operators in
the matter sector with $\gamma^{\ssc(a)}\cdot\gamma^{\ssc(a)}=2$.
Momentum conservation requires $\sum\gamma^{\ssc(a)}=0$ in the matter
sector, while the super-selection rules, for the gravity sector
given above, restrict the momenta to
\item{$z^{\ssc(1)}$:} $k=q$, $\tk=q+x$, $\al=-q$
\item{$z^{\ssc(2)}$:} $k=p$, $\tk=p+x$, $\al=-p$
\item{$z^{\ssc(3)}$:} $k=x/2-1/x$, $\tk=-1/x-x/2$, $\al=-x/2+1/x$
\item{$z^{\ssc(4)}$:} $k=-3x/2-1/x-p-q$, $\tk=-x/2-1/x-p-q$,
$\al=3x/2+1/x+p+q\ .$
\par\noindent
Again, one can explicitly verify that the result is \slc invariant,
and fixing $z^{\ssc(a)}=(\infty,1,z,0)$ yields
$$\cA=\int\dif^2z\ (1-z)^{\gamma^{\ssc(2)}\cdot\gamma^{\ssc(3)}}
\ z^{\gamma^{\ssc(4)}\cdot\gamma^{\ssc(3)}}
\ (1-\bar z)^{-1-x^2/2-px}\ {\bar z}^{x^2+x(p+q)}\ \ .$$
On the holomorphic side of this integral, one always has integral
exponents since $\gamma^{\ssc(a)}\cdot\gamma^{\ssc(b)}=-2,-1,0,1,2$,
while on the anti-holomorphic side, arbitrary exponents arise as $p$
and $q$ are varied. Thus in general the amplitude contains cuts
as are characteristic of correlators of the gravity dressings.
As $z$ approaches 1, one finds that for
$(\gamma^{\ssc(2)}\cdot\gamma^{\ssc(3)},p)=(-1,-x/2)$ or $(-2,1/x-x/2)$,
$\cA$ factorizes with the appropriate three-point amplitudes on
$|1-z|^{-1}$ or $|1-z|^{-2}$ singularities, respectively. Such
singularities are characteristic of poles in the Virasoro-Shapiro
amplitudes\cit{vs}. Similar results arise as $z$ approaches 0 or
$\infty$, as well.

\medskip
\centerline{7. Concluding remarks}
\smallskip

We have examined two-dimensional quantum gravity coupled to chiral
matter, and found that for fixed values of $c_\pm$, there is in fact a
family of theories labelled by the free parameter, $\xi$. One might
think of this result in analogy to the cosmological constant,
which arises as a free parameter in nonchiral gravity.
The effects of $\xi$ are far more intricate:
It does not
appear in the central charge of the combined
conformal and Lorentz induced action, eq.~\(cc), nor in the
partition function.
However, $\xi$ does affect critical exponents, see {\it e.g.},
eq.~\(ggaamm), the spectrum of discrete states, and
the positions of poles in amplitudes. Actually,
$\xi$ should be counted as only one of undetermined
couplings, which arise in association with the new interactions
 discussed in sect.~5.

We should note that in the light-cone analysis of chiral gravity by
ref.~\cite{oz}, it is concluded that $\xi$ is quantized to be $\xi=
ic_-/2$. Of course, we have found no evidence of such a quantization
condition in our analysis. This discrepancy may arise because the
stress tensor for the Lorentz field appears to be misidentified in
ref.~\cite{oz}.

One might think that $\xi$ should be fixed in analogy to the analysis
of the chiral gauge anomaly in two dimensions\cit{jac}, in which
no undetermined parameters arise.  While the formal expressions for the
chiral gauge and gravitational anomalies seem similar, it is important
to keep in mind that the spin connection drops out of the action for
fermions in 1+1 dimensions, so the determinant of interest possesses a
dependence on the spin connection only via quantum effects.
In the case of the
chiral gauge anomaly, the determinant depends explicitly
on one component of the connection---it therefore may be reasonable to
demand that the effective action also only involve the appropriate
component, thus fixing the analogue of $\xi.$

It is interesting to further investigate the effects of $\xi$ by calculating
the central charges of $\rho$ and $\theta$, separately,
$$\eqalign{
c_\rho=&\ 25-c_++\xi+{c_-^2\over4\xi}\cr
c_{\th,{\ssc R}} =&\ 1-c_--\xi-{c_-^2\over4\xi}\cr
c_{\th,{\ssc L}} =&\ 1+c_--\xi-{c_-^2\over4\xi}\ .\cr}
\eqno(central)$$
With $\xi\rightarrow\infty$, $c_\rho\rightarrow\infty$
as expected for
the semiclassical limit of the gravity theory. In this limit, the propagators
for both $\rho$ and $\theta$ vanish as can be seen from \(prop,newval).
As a result, exponential operators are not renormalized from the
classical expressions derived in accord with the analysis for \(doood)
and \(restr). For example, the area operator becomes simply
$\int\!\dif^2x\,\sqrt{\hat{g}}\,\ee^{2\rho}$. As well in this limit,
$\th\rightarrow\chi$, so that the discontinuities all occur in the
classical Lorentz phase. Thus one completely recovers the classical
world-sheet geometry in the limit $\xi\rightarrow\infty$.

{}From \(central), it would appear that $\xi\rightarrow0^+$ produces a limit
completely analogous to $\xi\rightarrow\infty$. In fact, this limit
must be taken with care in
the ($\rho$,$\th$) basis, as can be seen from the
fact that the kinetic term for $\th$ vanishes in \(rhoact). Of course, the
theory is perfectly well-defined at $\xi=0$, and no problems arise for
the original ($\rho$,$\chi$) basis, or the alternate diagonal basis
of ($\zeta$,$\chi$). While one still
has $\langle\rho(z,\bar z)\, \rho(w,\bar w)\rangle = 0$, so that
the area operator is unrenormalized in this limit,
exponentials involving both $\rho$ and $\chi$ do not
take their classical form. Further with this limit, it is
$\rho$ which becomes multivalued. Hence while $\xi\rightarrow0^+$
may allow for a semiclassical treatment of $\rho$, the interpretation
of the geometry is unclear in this limit.

A physically consistent
analysis required some restrictions on $\xi$ in the form of the
inequality \(jacket), but no barriers appeared for the matter theories.
Note as well that one can easily produce theories with
space-time tachyons, which do not yield unphysical critical exponents.
Thus at this level, the properties of space-time tachyons and
physical consistency on the world-sheet,
appear to be divorced in the present theory, in contrast to
nonchiral gravity\cit{seib2}. Clearly, the appearance of the Lorentz
field has drastic effects on the quantum theory of the world-sheet
geometry. The most pressing question would appear to be to
understand the complete space of physical states\cit{next}.

\bigskip
We thank J. Distler, S. Shatashvili and E. Witten for helpful comments.
R.C.M. gratefully acknowledges the hospitality of both the Aspen Center
for Physics, and the Institute for Theoretical Physics at UCSB, at various
stages of this work. R.C.M. was supported by NSERC of Canada, and Fonds
FCAR du Qu\'ebec. V.P. was supported by D.O.E. grant
DE-FG02-90ER40542, and thanks the Aspen Center for Physics for a
stimulating stay.

\references

\refis{next} In preparation

\refis{bas} F. Bastianelli, {\sl Phys. Lett.} {\bf 277B} (1992) 464

\refis{seib2} N. Seiberg, {\sl Prog. Theor. Phys. Suppl.}
{\bf 102} (1990) 319

\refis{fried} D. Friedan, in {\it Recent Advances in Field Theory and
Statistical Mechanics}, eds. J.-B. Zuber and R.~Stora, (Elsevier Science
Publishers, 1984).

\refis{vs} M.A. Virasoro, {\sl Phys. Rev.} {\bf 177} (1969) 2309;
J. Shapiro, {\sl Phys. Rev.} {\bf 179} (1969) 1345

\refis{shell} A.N. Schellekens, `Meromorphic $c=24$ conformal field
theories,' preprint CERN-TH.6478/92

\refis{pol} A.M. Polyakov, {\sl Mod. Phys. Lett.} {\bf A2} (1987) 893;
V.G. Knizhnik, A.M. Polyakov and A.B. Zamolodchikov, {\sl
Mod. Phys. Lett.} {\bf A3} (1988) 819

\refis{good} E.~Witten, {\sl Nucl. Phys.} {\bf B373} (1992) 187

\refis{mark} M.~Goulian and K.~Li, {\sl Phys. Rev. Lett} {\bf 66} (1991)
2051

\refis{dotf} Vl.S.~Dotsenko and V.~Fateev, {\sl Nucl. Phys.} {\bf B240}
(1984) 312; {\bf B251} (1985) 691

\refis{dp} See, for example,
E.~D'Hoker and D.H.~Phong, {\sl Rev. Mod. Phys.} {\bf 60}
(1988) 917

\refis{go} P.~Goddard and D.~Olive, in {\it Vertex Operators}, ed.
V.~Kac {\it et al.} (Springer, New York, 1983)

\refis{jps} J.-P.~Serre, {\it A Course in Arithmetic}, (Springer, New
York, 1973)

\refis{whit} See, for example, E.T. Whittaker and G.N. Watson, {\it
A Course in Modern Analysis} (Cambridge University Press, Cambridge,
1927)

\refis{jac} R. Jackiw, in {\it Relativity, Groups and Topology II},
eds. B.S.~DeWitt and R.~Stora, (North-Holland, Amsterdam, 1984).

\refis{ddk} F. David, {\sl Mod. Phys. Lett.} {\bf A3} (1988) 1651;
J. Distler and H. Kawai, {\sl Nucl. Phys.} {\bf B321} (1989) 509

\refis{rob} R.C. Myers, {\sl Phys. Lett.} {\bf 199B} (1987) 371

\refis{oz} Y. Oz, J. Pawelczyk and S. Yankielowicz, {\sl Phys. Lett.}
{\bf B249} (1990) 417; {\sl Nucl. Phys.} {\bf B363} (1991) 555

\refis{mmdk} N.E. Mavromatos and J.L. Miramontes, {\sl Mod. Phys. Lett.}
{\bf A4} (1989) 1847; E. D'Hoker and P.S. Kurzepa, {\sl Mod. Phys. Lett.}
{\bf A5} (1990) 1411

\refis{eric} E. D'Hoker, {\sl Mod. Phys. Lett.} {\bf A6} (1991) 745

\refis{leu} H. Leutwyler, {\sl Phys. Lett.} {\bf 153B} (1985) 65

\refis{bank} T. Banks, D. Nemeschansky and A. Sen, {\sl Nucl.Phys.}
{\bf B277} (1986) 67

\endreferences
\end